\documentclass[]{aastex}

\usepackage{emulateapj5}
\usepackage{onecolfloat}
\usepackage{graphicx} 
\usepackage{fancyheadings} 
\usepackage{ulem}
\usepackage{rotating}
\usepackage{lscape}

\newcommand{\ndla}{12}

\newcommand{\etal}{et al.\ }

\newcommand{\lya}{Ly$\alpha$ }

\newcommand{\kms}{km~s$^{-1}$ }
\newcommand{\cm}[1]{\, {\rm cm^{#1}}}
\newcommand{\N}[1]{{N({\rm #1})}}

\newcommand{\mkms}{{\rm \; km\;s^{-1}}}
\newcommand{\tskip}{\tablevspace{1pt}}

\begin{document}

\twocolumn[%
\submitted{Accepted to the Astrophysical Journal Supplements: May 9, 2003}
\title{New Damped \lya Metallicities from ESI Spectroscopy
of Five Palomar Sky Survey Quasars}

\author{ JASON X. PROCHASKA}
\affil{UCO/Lick Observatory\\ 
University of California, Santa Cruz; Santa Cruz, CA 95064}
\author{SANDRA CASTRO\altaffilmark{1,2} \& 
S. G. DJORGOVSKI\altaffilmark{1}} 
\affil{Palomar Observatory, California Institute of Technology\\
MS 105-24; Pasadena, CA 91125}

\begin{abstract} 

This paper presents chemical abundance measurements for \ndla\ new 
$z>3$ damped
\lya systems discovered toward five quasars from the Palomar Sky Survey.
We determine H\,I column densities from profile fits to the observed
damped \lya profiles and measure ionic
column densities and limits for all observed metal-line transitions.  
This dataset, acquired with the Echellette Spectrograph and Imager 
on the Keck~II telescope, adds to the rapidly growing database of damped \lya 
abundances.  It will impact studies of chemical evolution in
the early universe and help identify candidates for detailed follow-up
observations with echelle spectrographs.  We report the discovery of
the first quasar sightline with four cosmologically distinct damped \lya 
systems.

\keywords{galaxies: abundances --- 
galaxies: chemical evolution --- quasars : absorption lines ---
nucleosynthesis}

\end{abstract}
]

\altaffiltext{1}{Visiting Astronomer, W.M. Keck Telescope.
The Keck Observatory is a joint facility of the University
of California, California Institute of Technology, and NASA.}
\altaffiltext{2}{Current address: Infrared Processing and
Analysis Center, 100-22, California Institute of Technology,
Pasadena, CA, 91125}

\pagestyle{fancyplain}
\lhead[\fancyplain{}{\thepage}]{\fancyplain{}{PROCHASKA ET AL.}}
\rhead[\fancyplain{}{New Damped \lya Metallicities from ESI Spectroscopy
of Five Palomar Sky Survey Quasars}]{\fancyplain{}{\thepage}}
\setlength{\headrulewidth=0pt}
\cfoot{}

\section{INTRODUCTION}
\label{sec-intro}

Studies of absorption line systems in the spectra of high-redshift QSOs
continue to provide valuable insights into the early chemical evolution of
galaxies and IGM.  One sample of bright ($r \lesssim 19.5$ mag), high-redshift
($z \sim 3.9 - 4.6$) QSOs is the sample discovered in Digital Palomar
Observatory Sky Survey (DPOSS; Djorgovski \etal 1999, and in prep.).
The QSOs were discovered using now standard color selection techniques;
for more details, see, e.g., Kennefick \etal 1995a, 1995b, or Djorgovski
\etal 2001.  The sample (commonly designated as PSS, for Palomar Sky Survey),
including many as yet unpublished objects, is available on 
line\footnote{\tt http://www.astro.caltech.edu/$\sim$george/z4.qsos}.

Relatively high--resolution, high-S/N 
($R \approx 4500$, S/N~$> 15$ per 10km/s pixel)
spectra of nearly all PSS QSOs and a number of
additional bright, high-$z$ QSOs from other surveys, were obtained using the
ESI instrument (Sheinis \etal 2002) at the Keck II 10-m telescope in the course
of a survey for high column density absorbers.  The results of this survey
will be presented in future papers (Castro \etal 2003, Djorgovski
\etal, in prep.).  
This paper serves as a companion to the ESI/Keck~II Damped \lya Abundance
Database comprised by Prochaska et al.\ (2003; hereafter P03).
Here we present the chemical abundance measurements from the
PSS damped \lya sample avoiding repetition with the results presented in P03.
We analyse 12 damped \lya systems toward 5 quasars.  This incidence of damped
\lya systems is higher than a random sample because we have restricted the
analysis to sightlines showing at least one DLA.  Even still, the incidence
is higher than expected from the number density of DLA in the literature
\citep[e.g.][]{storrie00} perhaps owing to the higher resolution of our
spectra but more likely small number statistics.
We present the relevant figures and tables related to our abundance analysis
and withhold extensive analysis to future papers.
This paper is outlined as follows.
We describe the observations and data reduction routines in $\S$~2.  
The individual damped \lya systems are presented in $\S$~3 
and $\S$~4 provides a brief summary.

\begin{table}[ht]\footnotesize
\begin{center}
\caption{ {\sc JOURNAL OF OBSERVATIONS\label{tab:obs}}}
\begin{tabular}{lccccl}
\tableline
\tableline
Quasar
& $R$
& $z_{em}$
& Date
& Exp (s)
& Arcs \\
\tableline
PSS0007+2417 & 18.7 & 4.05 & 06sep00 & 3600 & CuAr \\
PSS1535+2943 & 18.9 & 3.99 & 15may00 & 5400 & CuAr+Xe \\
PSS1715+3809 & 18.6 & 4.52 & 03sep00 & 3600 & CuAr \\
PSS1802+5616 & 19.2 & 4.18 & 05sep00 & 5400 & CuAr \\
PSS2315+0921 & 19.5 & 4.30 & 05sep00 & 5400 & CuAr \\
\tableline
\end{tabular}
\end{center}
\end{table}

\begin{figure*}[ht]
\begin{center}
\includegraphics[height=5.5in, width=4.0in,angle=90]{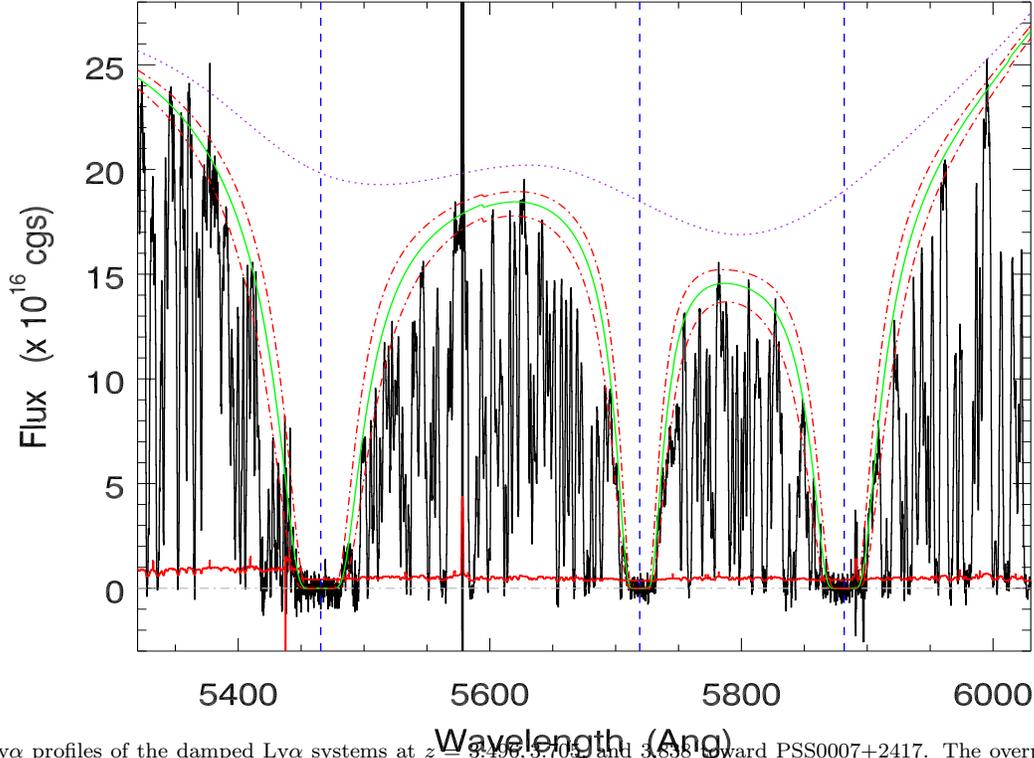}
\figcaption{\lya profiles of the damped \lya systems at $z=3.496, 3.705$,
and 3.838 toward PSS0007+2417.
The overplotted solid line and accompanying
dash-dot lines trace the best fit solutions corresponding to 
$\log \N{HI} = 21.10^{+0.10}_{-0.10}, 20.55^{+0.15}_{-0.15}$
and 20.85$^{+0.15}_{-0.15}$.  
The solution for all three damped \lya systems are reasonably well
constrained by both the wings and cores of the profiles.  One notes 
that the predicted continuum exhibits significant variation, primarily
due to the presence of the \lya and O\,VI emission peaks.  
Because the continuum is more poorly constrained from 5600--6000\AA,
we report 0.15~dex uncertainties for the two higher redshift DLA.
\label{fig:pss0007_lya}}
\end{center}
\end{figure*}

\begin{figure}[ht]
\begin{center}
\includegraphics[height=6.1in, width=3.9in]{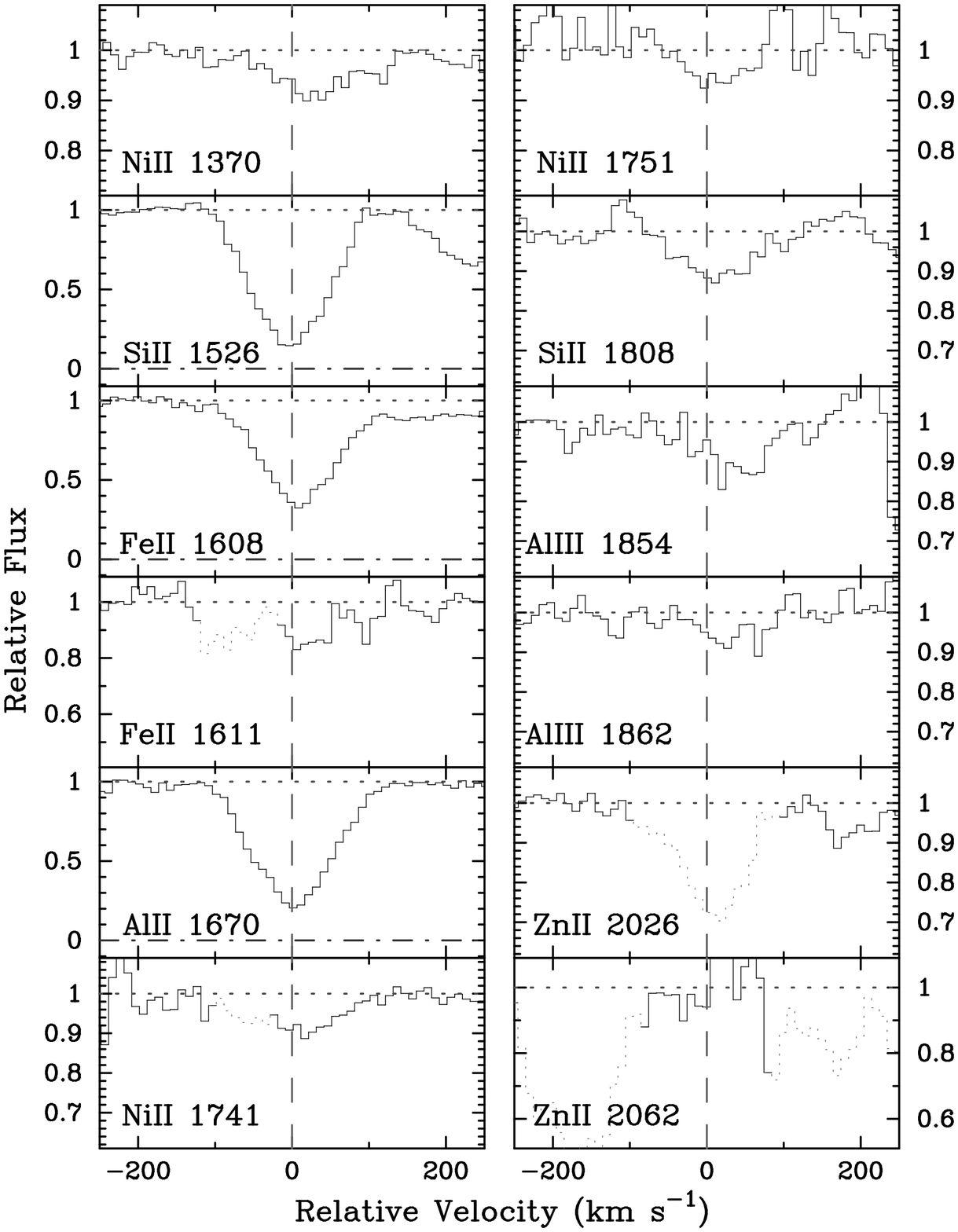}
\figcaption{Velocity plot of the metal-line transitions for the 
damped \lya system at $z = 3.496$ toward PSS0007+2417.
The vertical line at $v=0$ corresponds to $z = 3.4962$.  
\label{fig:pss0007A_mtl}}
\end{center}
\end{figure}

\begin{figure}[ht]
\begin{center}
\includegraphics[height=6.1in, width=3.9in]{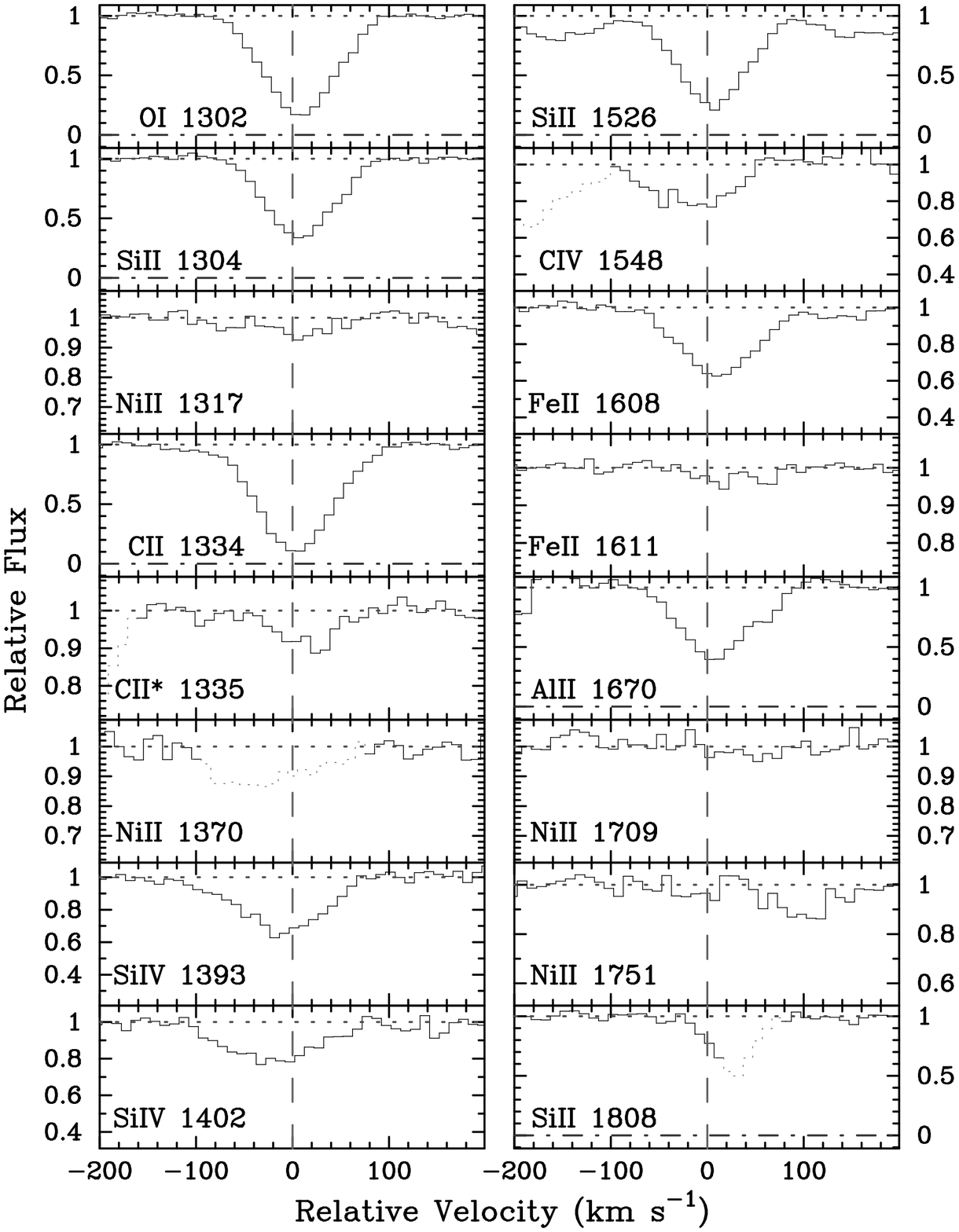}
\figcaption{Velocity plot of the metal-line transitions for the 
damped \lya system at $z = 3.705$ toward PSS0007+2417.
The vertical line at $v=0$ corresponds to $z = 3.70454$.  
\label{fig:pss0007B_mtl}}
\end{center}
\end{figure}

\begin{figure}[ht]
\begin{center}
\includegraphics[height=6.1in, width=3.9in]{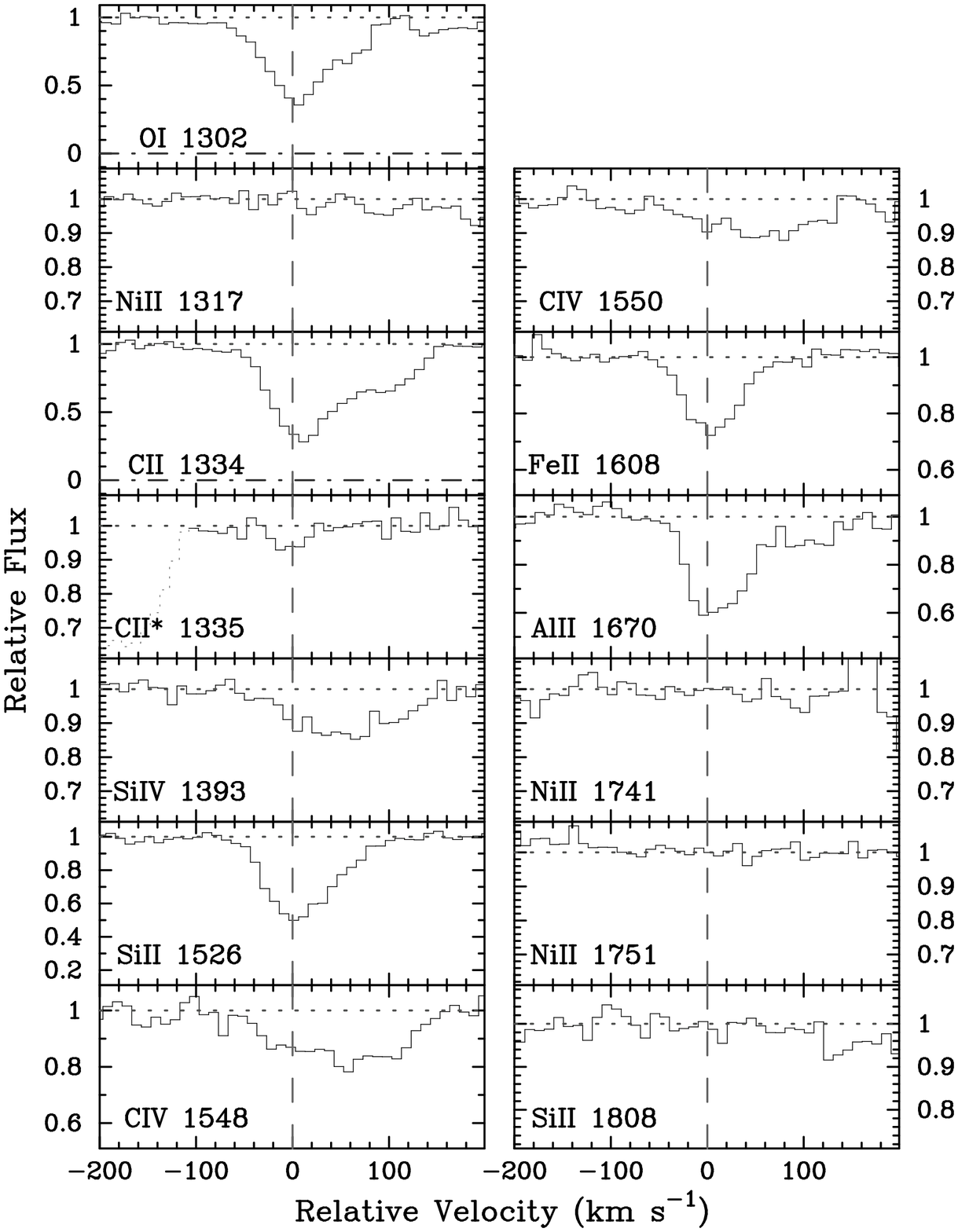}
\figcaption{Velocity plot of the metal-line transitions for the 
damped \lya system at $z = 3.838$ toward PSS0007+2417.
The vertical line at $v=0$ corresponds to $z = 3.83823$.  
\label{fig:pss0007C_mtl}}
\end{center}
\end{figure}

\begin{table}[ht]\footnotesize
\begin{center}
\caption{ {\sc
IONIC COLUMN DENSITIES: PSS0007+2417, $z = 3.496$ \label{tab:PSS0007+2417_3.496}}}
\begin{tabular}{lcccc}
\tableline
\tableline
Ion & $\lambda$ & AODM & $N_{\rm adopt}$ & [X/H] \\
\tableline
Al II &1670.8&$>13.249$&$>13.249$&$>-2.341$\\  
Al III&1854.7&$12.694 \pm  0.052$\\  
Al III&1862.8&$12.696 \pm  0.086$\\  
Si II &1526.7&$>14.520$&$15.077 \pm  0.040$&$-1.583 \pm  0.108$\\  
Si II &1808.0&$15.077 \pm  0.040$\\  
Fe II &1608.5&$>14.630$&$>14.630$&$>-1.970$\\  
Fe II &1611.2&$<15.262$\\  
Ni II &1370.1&$13.518 \pm  0.040$&$13.529 \pm  0.036$&$-1.821 \pm  0.106$\\  
Ni II &1741.6&$<13.842$\\  
Ni II &1751.9&$13.591 \pm  0.081$\\  
Zn II &2062.7&$<12.390$&$<12.390$&$<-1.380$\\  
\tableline
\end{tabular}
\end{center}
\end{table}
\begin{table}[ht]\footnotesize
\begin{center}
\caption{ {\sc
IONIC COLUMN DENSITIES: PSS0007+2417, $z = 3.705$ \label{tab:PSS0007+2417_3.705}}}
\begin{tabular}{lcccc}
\tableline
\tableline
Ion & $\lambda$ & AODM & $N_{\rm adopt}$ & [X/H] \\
\tableline
C  II &1334.5&$>14.530$&$>14.530$&$>-2.610$\\  
C  II*&1335.7&$13.242 \pm  0.061$\\  
C  IV &1548.2&$13.460 \pm  0.031$\\  
O  I  &1302.2&$>14.855$&$>14.855$&$>-2.435$\\  
Al II &1670.8&$>12.895$&$>12.895$&$>-2.145$\\  
Si II &1304.4&$>14.372$&$>14.372$&$>-1.738$\\  
Si II &1526.7&$>14.291$\\  
Si II &1808.0&$<14.854$\\  
Si IV &1393.8&$13.329 \pm  0.019$\\  
Si IV &1402.8&$13.433 \pm  0.032$\\  
Fe II &1608.5&$>14.211$&$>14.211$&$>-1.839$\\  
Fe II &1611.2&$<14.603$\\  
Ni II &1370.1&$<13.612$&$<13.197$&$<-1.603$\\  
Ni II &1709.6&$<13.311$\\  
Ni II &1751.9&$<13.507$\\  
\tableline
\end{tabular}
\end{center}
\end{table}
\begin{table}[ht]\footnotesize
\begin{center}
\caption{ {\sc
IONIC COLUMN DENSITIES: PSS0007+2417, $z = 3.838$ \label{tab:PSS0007+2417_3.838}}}
\begin{tabular}{lcccc}
\tableline
\tableline
Ion & $\lambda$ & AODM & $N_{\rm adopt}$ & [X/H] \\
\tableline
C  II &1334.5&$>14.394$&$>14.394$&$>-3.046$\\  
C  II*&1335.7&$<12.891$\\  
C  IV &1548.2&$13.579 \pm  0.024$\\  
C  IV &1550.8&$13.614 \pm  0.037$\\  
O  I  &1302.2&$>14.643$&$>14.643$&$>-2.947$\\  
Al II &1670.8&$12.699 \pm  0.018$&$12.699 \pm  0.019$&$-2.641 \pm  0.151$\\  
Si II &1526.7&$>14.006$&$>14.006$&$>-2.404$\\  
Si II &1808.0&$<14.441$\\  
Si IV &1393.8&$12.962 \pm  0.035$\\  
Fe II &1608.5&$13.906 \pm  0.031$&$13.906 \pm  0.031$&$-2.444 \pm  0.153$\\  
Ni II &1317.2&$<13.042$&$<13.042$&$<-2.058$\\  
Ni II &1741.6&$<13.234$\\  
Ni II &1751.9&$<13.399$\\  
\tableline
\end{tabular}
\end{center}
\end{table}

\section{OBSERVATIONS, DATA REDUCTION AND ANALYSIS}
\label{sec:redux}

With the commissioning of the Echellette Spectrograph and Imager, 
the DPOSS team (PI: Djorgovski) 
initiated an observing campaign to survey $z>3$ damped \lya systems.
A full account of this survey will be presented in a future paper.
In Table~\ref{tab:obs} we present a journal of the observations
for the background quasars of the damped \lya systems presented in this paper.  
Column~1 gives the name,
column~2 is the apparent $R$ magnitude of the quasar, 
column~3 is the emission redshift,
column~4 gives the date of the observation,
column~5 is the exposure time, and
column~6 lists the arc calibrations available for data reduction.
To avoid repetition, we have restricted the sample analysed here
to the sub-set which
were not analysed by P03.  The P03 survey has comparable S/N data but 
higher resolution ($R \approx 9000$) and therefore more accurate abundance
measurements.

Data reduction was performed using the ESI/IDL package built by 
Prochaska (P03).  In contrast to the P03 database, the majority of 
observations presented here were calibrated with the CuAr lamps only.
This limits the spectral coverage to $\lambda < 9000$\AA\ due to a
steep drop in transmissivity of the fibers which feed the CuAr lamps
to the spectrograph.  The other major difference in the datasets is that
all of the data were taken with the 1.0$''$ slit (the majority of data
in P03 used the 0.5$''$ slit).  Both the lower resolution and the 
limited arc calibration frames imply poorer sky line subtraction 
at $\lambda > 6000$\AA.  Nevertheless, the spectra provide multiple
metal-line transitions for most of the damped \lya systems considered
here and reasonably accurate abundance measurements.
In the future, however, we may implement a
reduction scheme using the night sky lines to better calibrate the reddest
orders of the spectra.

Measurements of H\,I column densities $\N{HI}$ and ionic column densities
were determined using the techniques described in P03.  
Specifically, we performed Voigt profile fits to the \lya transitions
and utilized the apparent optical depth method \citep{sav91}
to evaluate column densities of unsaturated, unblended metal-line 
transitions.  Because the \lya profiles extend many angstroms of spectra,
the lower resolution of this dataset in comparison with P03
implies a minimal difference in
$\N{HI}$ precision.  
While our H\,I analysis is not based on a $\chi^2$ minimization technique,
we report conservative statistical errors which we contend correspond to 
95$\%$c.l. (see P03 for a complete discussion).
For the ionic column densities, we have taken
greater care to account for  the effects of line-saturation.  In general, we have
been very conservative in our analysis, reporting lower limits in
those cases where line-saturation is possible.
Generally, this included all profiles where the peak optical depth
of any 10~\kms\ pixel exceeds 0.5 (i.e.\ normalized flux less than 60$\%$
the quasar continuum).
Finally although we report formal statistical errors from our
analysis of the ionic column densities,  we caution that the systematic
effects of line-saturation and continuum placement will lend larger
uncertainties in most cases.  We recommend that the reader assume 1$\sigma$
uncertainties of at least 0.1~dex for all measurements.

Throughout the paper, we adopt the wavelengths
and oscillator strengths presented in Table~2 of P03.
We also adopt solar meteoritic abundances from \cite{grvss96} and 
\cite{holweger01} for the elemental abundances listed in the paper.

\begin{figure}[ht]
\begin{center}
\includegraphics[height=3.6in, width=2.8in,angle=90]{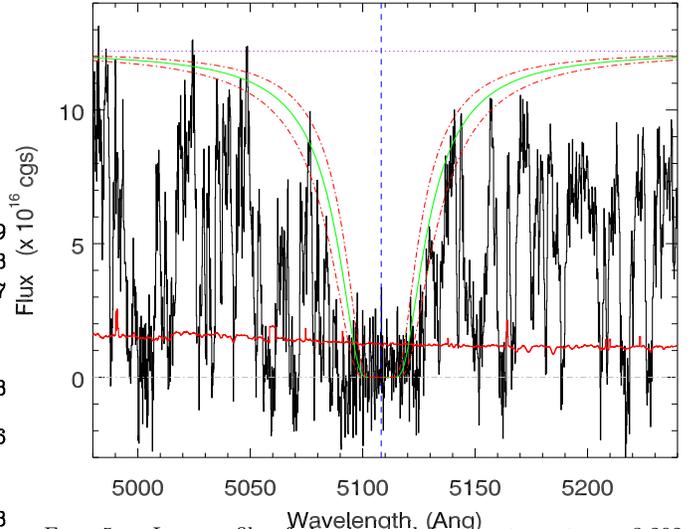}
\figcaption{\lya profile of the damped \lya system at $z=3.202$
toward PSS1535+2943.
The overplotted solid line and accompanying
dash-dot lines trace the best fit solution corresponding to 
$\log \N{HI} = 20.65^{+0.15}_{-0.15}$.
Although the quasar continuum is well determined across this region of the 
spectrum, the core of this \lya profile is not well sampled and we assign
an uncertainty of 0.15~dex.
\label{fig:pss1535A_lya}}
\end{center}
\end{figure}

\begin{figure}[ht]
\begin{center}
\includegraphics[height=3.6in, width=2.8in,angle=90]{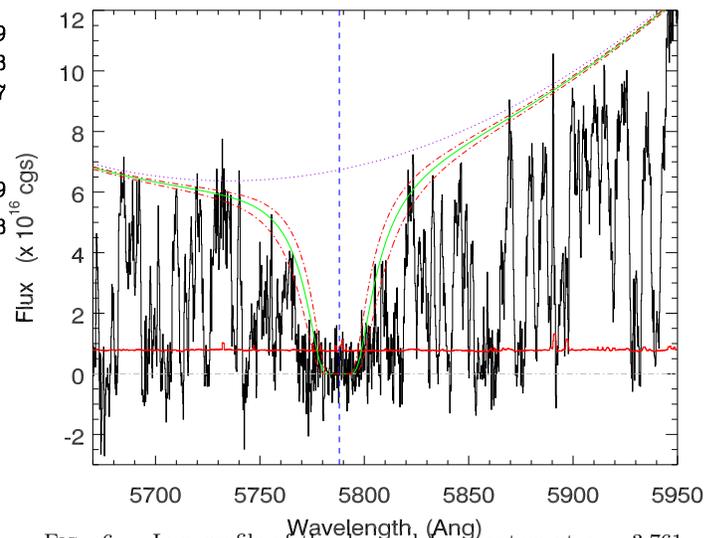}
\figcaption{\lya profile of the damped \lya system at $z=3.761$
toward PSS1535+2943.
The overplotted solid line and accompanying
dash-dot lines trace the best fit solution corresponding to 
$\log \N{HI} = 20.40^{+0.15}_{-0.15}$.
\label{fig:pss1535B_lya}}
\end{center}
\end{figure}

\begin{figure}[ht]
\begin{center}
\includegraphics[height=6.1in, width=3.9in]{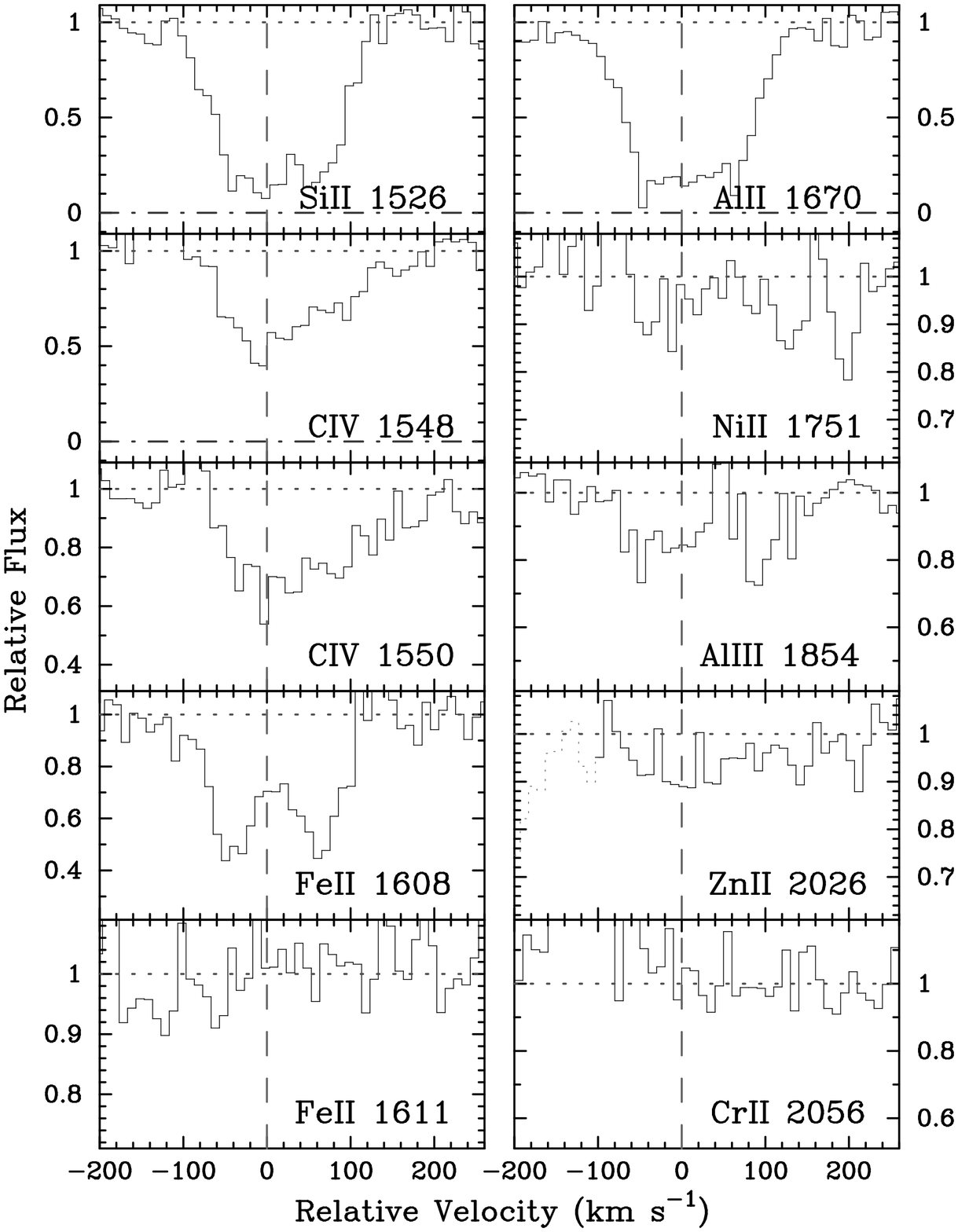}
\figcaption{Velocity plot of the metal-line transitions for the 
damped \lya system at $z = 3.202$ toward PSS1535+2943.
The vertical line at $v=0$ corresponds to $z = 3.2020$.  
\label{fig:pss1535A_mtl}}
\end{center}
\end{figure}

\begin{figure}[ht]
\begin{center}
\includegraphics[height=6.1in, width=3.9in]{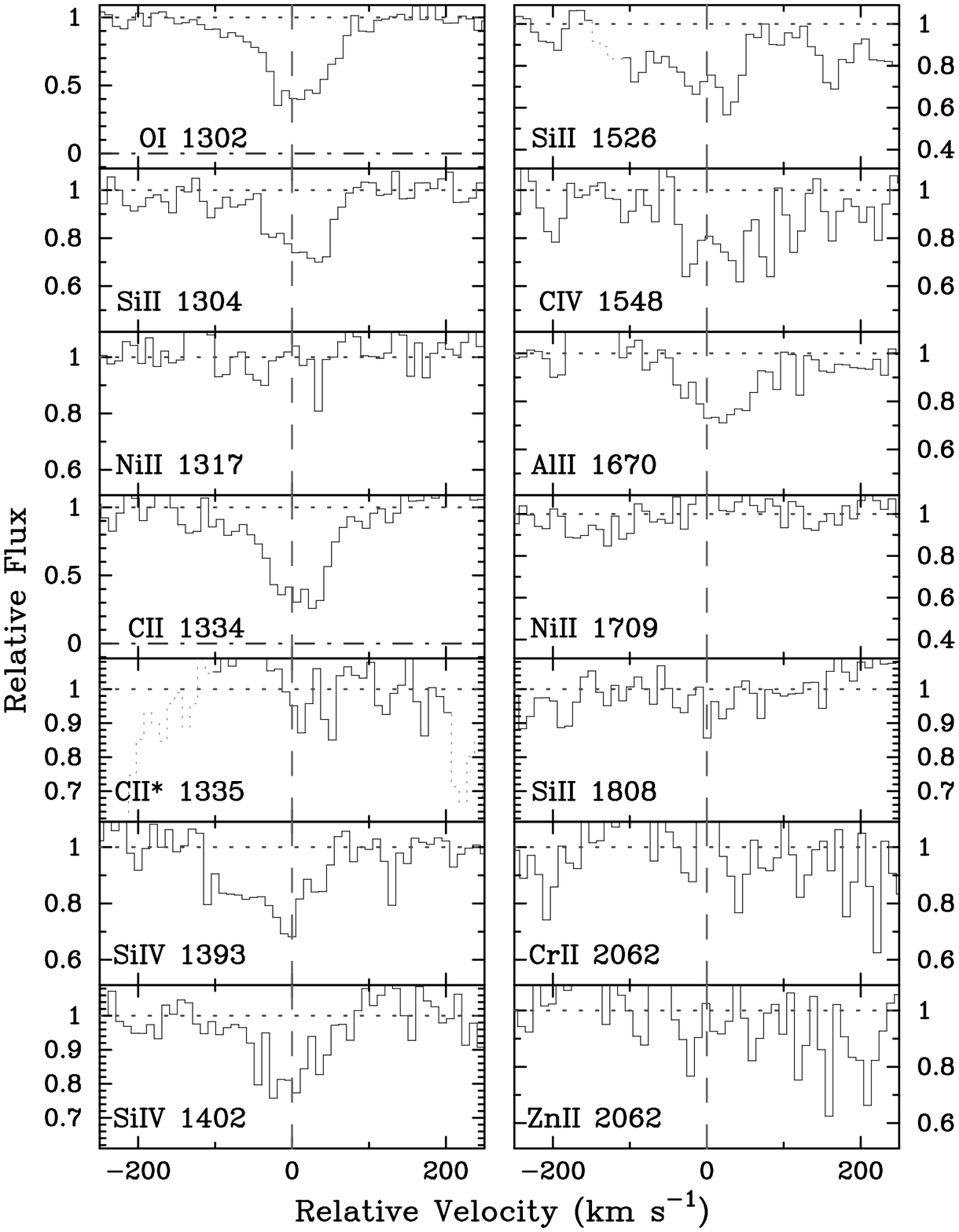}
\figcaption{Velocity plot of the metal-line transitions for the 
damped \lya system at $z = 3.761$ toward PSS1535+2943.
The vertical line at $v=0$ corresponds to $z = 3.7612$.  
\label{fig:pss1535B_mtl}}
\end{center}
\end{figure}

\begin{table}[ht]\footnotesize
\begin{center}
\caption{ {\sc
IONIC COLUMN DENSITIES: PSS1535+2943, $z = 3.202$ \label{tab:PSS1535+2943_3.202}}}
\begin{tabular}{lcccc}
\tableline
\tableline
Ion & $\lambda$ & AODM & $N_{\rm adopt}$ & [X/H] \\
\tableline
C  IV &1548.2&$<14.085$\\  
C  IV &1550.8&$14.232 \pm  0.040$\\  
Al II &1670.8&$>13.560$&$>13.560$&$>-1.580$\\  
Al III&1854.7&$12.848 \pm  0.095$\\  
Si II &1526.7&$>14.742$&$>14.742$&$>-1.468$\\  
Cr II &2056.3&$<13.189$&$<13.189$&$<-1.131$\\  
Fe II &1608.5&$>14.603$&$>14.603$&$>-1.547$\\  
Fe II &1611.2&$<15.198$\\  
Ni II &1751.9&$<13.972$&$<13.972$&$<-0.928$\\  
Zn II &2026.1&$<12.531$&$<12.531$&$<-0.789$\\  
\tableline
\end{tabular}
\end{center}
\end{table}

\begin{table}[ht]\footnotesize
\begin{center}
\caption{ {\sc
IONIC COLUMN DENSITIES: PSS1535+2943, $z = 3.761$ \label{tab:PSS1535+2943_3.761}}}
\begin{tabular}{lcccc}
\tableline
\tableline
Ion & $\lambda$ & AODM & $N_{\rm adopt}$ & [X/H] \\
\tableline
C  II &1334.5&$>14.369$&$>14.369$&$>-2.621$\\  
C  II*&1335.7&$<13.291$\\  
C  IV &1548.2&$13.712 \pm  0.079$\\  
O  I  &1302.2&$>14.711$&$>14.711$&$>-2.429$\\  
Al II &1670.8&$12.563 \pm  0.051$&$12.563 \pm  0.051$&$-2.327 \pm  0.158$\\  
Si II &1304.4&$13.940 \pm  0.048$&$13.940 \pm  0.049$&$-2.020 \pm  0.158$\\  
Si II &1526.7&$13.970 \pm  0.040$\\  
Si II &1808.0&$<14.819$\\  
Si IV &1393.8&$13.242 \pm  0.064$\\  
Si IV &1402.8&$13.356 \pm  0.085$\\  
Ni II &1317.2&$<13.453$&$<13.453$&$<-1.197$\\  
Ni II &1709.6&$<13.690$\\  
\tableline
\end{tabular}
\end{center}
\end{table}

\section{INDIVIDUAL SYSTEMS}

In this section we present the metal-line and \lya profiles of the
DLA and tabulate the measured ionic column densities.  
For the metal-line transitions, we integrated unblended velocity profiles
over a velocity region including all significant absorption.  The
exact values can be obtained from the authors upon request.
Regarding the figures, $v=0$ \kms\ is arbitrarily defined and blends from
coincident absorption lines are indicated by dotted lines.
For the low-ion transitions -- 
the dominant ionic transition of a given element in 
an H\,I region -- we tabulate $N_{adopt}$ by taking the weighted
mean (or limit) of all measured transitions.  For these cases, we also
convert $N_{adopt}$ into an elemental abundance [X/H] assuming no
ionization corrections and adopting solar meteoritic abundances.

\begin{figure}[ht]
\begin{center}
\includegraphics[height=3.6in, width=2.8in,angle=90]{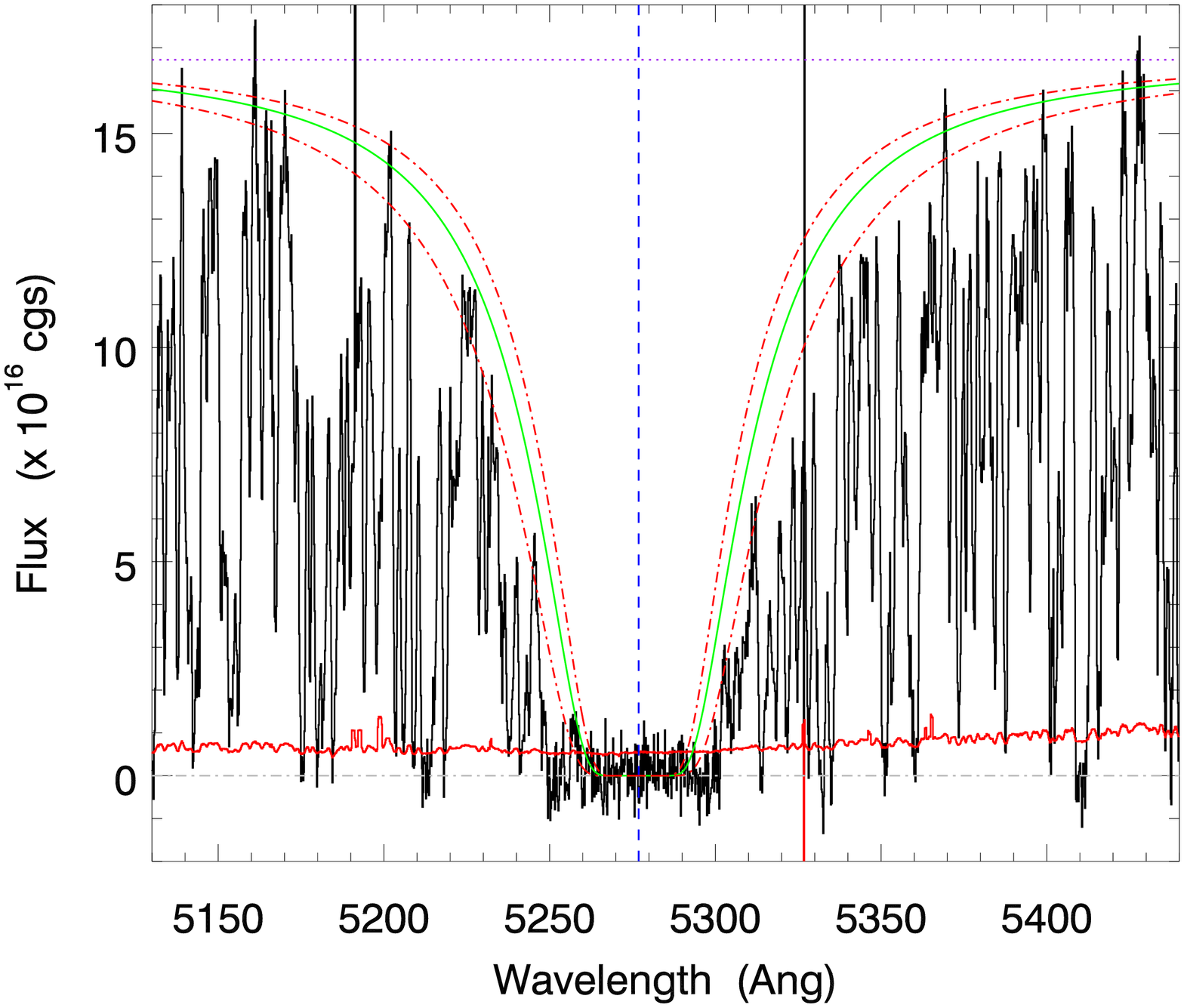}
\figcaption{\lya profile of the damped \lya system at $z=3.341$
toward PSS1715+3809.
The overplotted solid line and accompanying
dash-dot lines trace the best fit solution corresponding to 
$\log \N{HI} = 21.05^{+0.15}_{-0.10}$.
Although it may appear from the figure that we have underestimated
the $\N{HI}$ value, the flux at $\lambda \approx 5260$\AA\ is significant
and the continuum is well constrained.  Together, these data imply
the value we have adopted. 
\label{fig:pss1715_lya}}
\end{center}
\end{figure}

\begin{figure}[ht]
\begin{center}
\includegraphics[height=6.1in, width=3.9in]{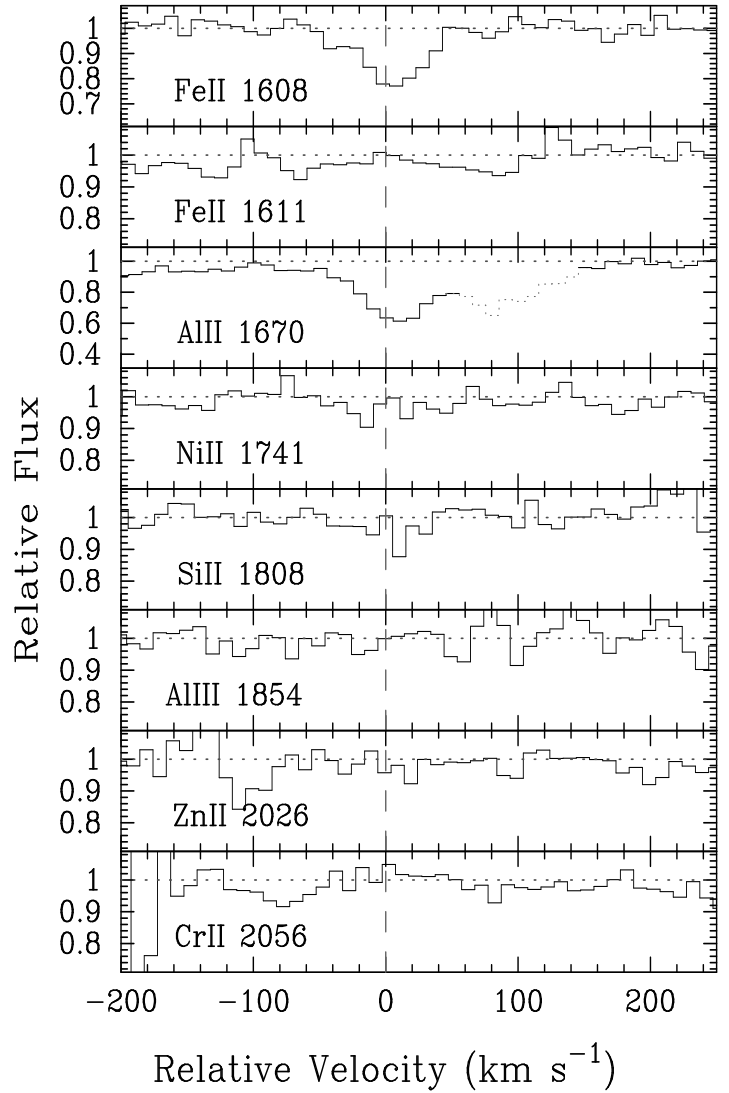}
\figcaption{Velocity plot of the metal-line transitions for the 
damped \lya system at $z = 3.341$ toward PSS1715+3809.
The vertical line at $v=0$ corresponds to $z = 3.3407$.  
\label{fig:pss1715_mtl}}
\end{center}
\end{figure}

\begin{table}[ht]\footnotesize
\begin{center}
\caption{ {\sc
IONIC COLUMN DENSITIES: PSS1715+3809, $z = 3.341$ \label{tab:PSS1715+3809_3.341}}}
\begin{tabular}{lcccc}
\tableline
\tableline
Ion & $\lambda$ & AODM & $N_{\rm adopt}$ & [X/H] \\
\tableline
Al II &1670.8&$12.551 \pm  0.021$&$12.551 \pm  0.021$&$-2.989 \pm  0.127$\\  
Al III&1854.7&$<12.325$\\  
Si II &1808.0&$<14.488$&$<14.488$&$<-2.122$\\  
Cr II &2056.3&$<12.773$&$<12.773$&$<-1.947$\\  
Fe II &1608.5&$13.743 \pm  0.039$&$13.743 \pm  0.039$&$-2.807 \pm  0.131$\\  
Fe II &1611.2&$<14.699$\\  
Ni II &1741.6&$<13.294$&$<13.294$&$<-2.006$\\  
Zn II &2026.1&$<12.113$&$<12.113$&$<-1.607$\\  
\tableline
\end{tabular}
\end{center}
\end{table}

\subsection{PSS0007+2417, $z = 3.496,3.705,3.838$ \label{subsec:PSS0007+2417}}

This quasar exhibits three damped \lya systems along its sightline
whose \lya profiles are shown together in Figure~\ref{fig:pss0007_lya}.
The $\N{HI}$ values are reasonably well constrained although the proximity
of the DLA and the presence of the O\,VI and \lya emission peaks do 
complicate the continuum placement.  All three DLA exhibit a significant
number of metal-line transitions 
(Figures~\ref{fig:pss0007A_mtl}-\ref{fig:pss0007C_mtl}) and 
have reasonably well determined, low metallicities.
The ionic column densities and elemental abundances are listed in
Tables~\ref{tab:PSS0007+2417_3.496}-\ref{tab:PSS0007+2417_3.838}.

\begin{figure}[ht]
\begin{center}
\includegraphics[height=3.6in, width=2.8in,angle=90]{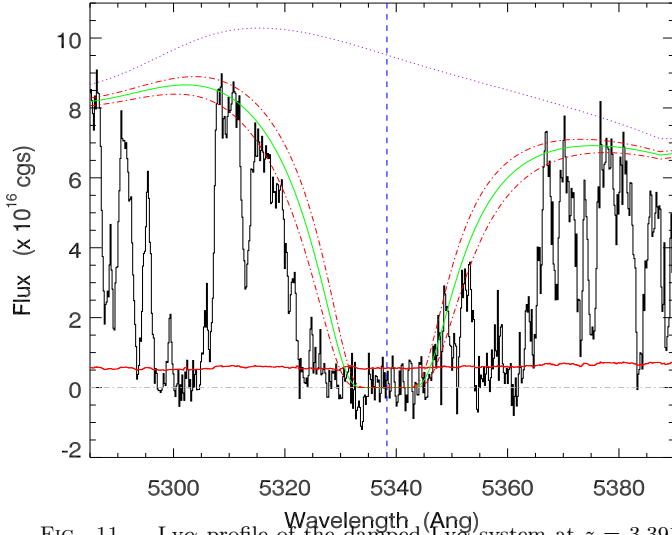}
\figcaption{\lya profile of the damped \lya system at $z=3.391$
toward PSS1802+5616.
The overplotted solid line and accompanying
dash-dot lines trace the best fit solution corresponding to 
$\log \N{HI} = 20.30^{+0.10}_{-0.10}$.
This damped system just satisfies the DLA criterion of 
$2 \times 10^{20} \cm{-2}$.  Although the fit to the \lya profile
is complicated by the O\,VI emission peak, the $\N{HI}$ value is well 
constrained by the wings of the profile.
\label{fig:pss1802A_lya}}
\end{center}
\end{figure}

\begin{figure}[ht]
\begin{center}
\includegraphics[height=3.6in, width=2.8in,angle=90]{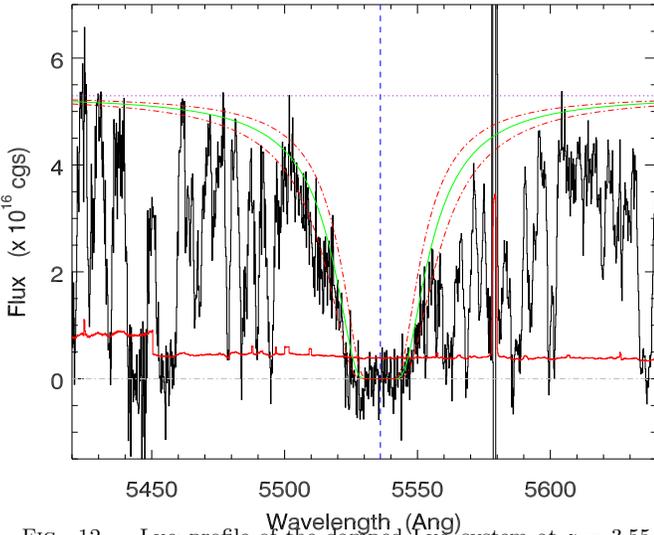}
\figcaption{\lya profile of the damped \lya system at $z=3.554$
toward PSS1802+5616.
The overplotted solid line and accompanying
dash-dot lines trace the best fit solution corresponding to 
$\log \N{HI} = 20.50^{+0.10}_{-0.10}$.
\label{fig:pss1802B_lya}}
\end{center}
\end{figure}

\begin{figure}[ht]
\begin{center}
\includegraphics[height=3.6in, width=2.8in,angle=90]{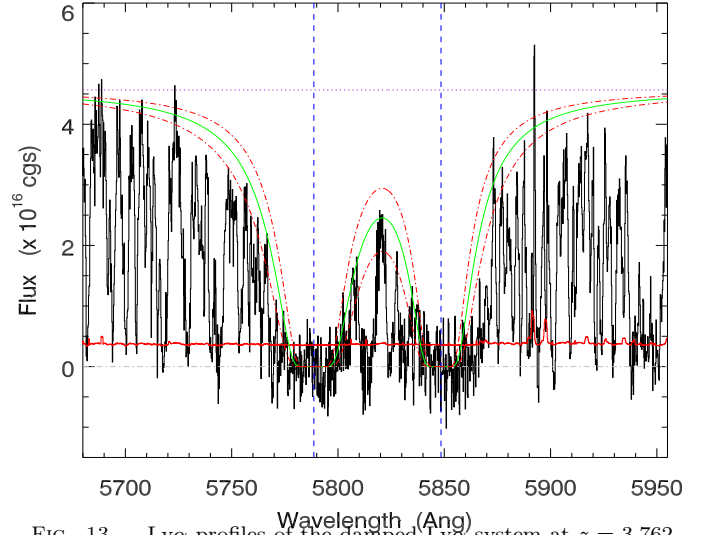}
\figcaption{\lya profiles of the damped \lya system at $z=3.762$
and 3.811 toward PSS1802+5616.
The overplotted solid line and accompanying
dash-dot lines trace the best fit solution corresponding to 
$\log \N{HI} = 20.55^{+0.15}_{-0.15}$ and 
20.35$^{+0.20}_{-0.20}$ respectively.
The proximity of the two DLA complicates the solution and we
report $\N{HI}$ values with lower precision.  In particular,
there is little data outside of $\lambda \approx 5820$\AA\
constraining the $z=3.811$ DLA.
\label{fig:pss1802C_lya}}
\end{center}
\end{figure}

\begin{figure}[ht]
\begin{center}
\includegraphics[height=6.1in, width=3.9in]{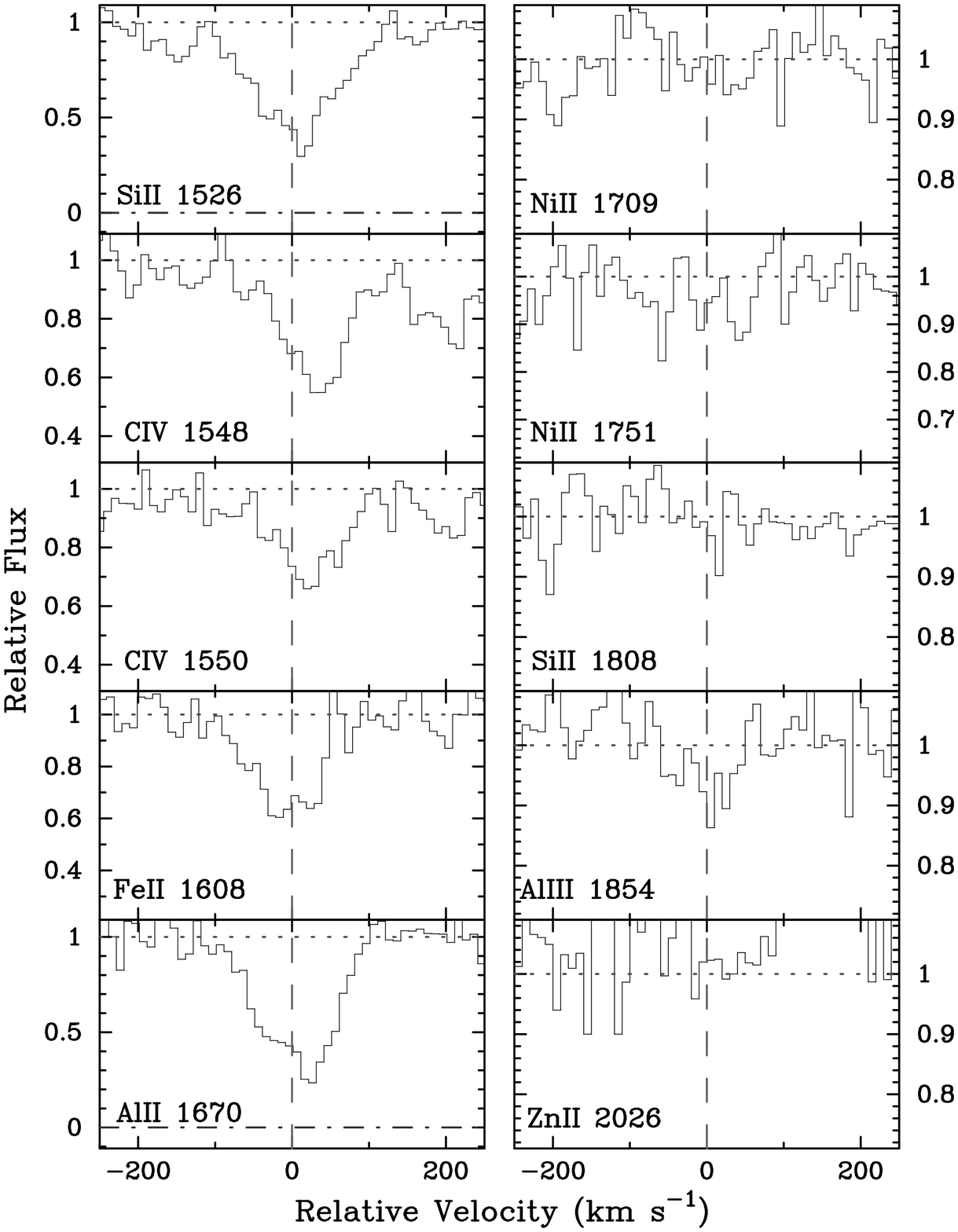}
\figcaption{Velocity plot of the metal-line transitions for the 
damped \lya system at $z = 3.391$ toward PSS1802+5616.
The vertical line at $v=0$ corresponds to $z = 3.39126$.  
\label{fig:pss1802A_mtl}}
\end{center}
\end{figure}

\begin{figure}[ht]
\begin{center}
\includegraphics[height=6.1in, width=3.9in]{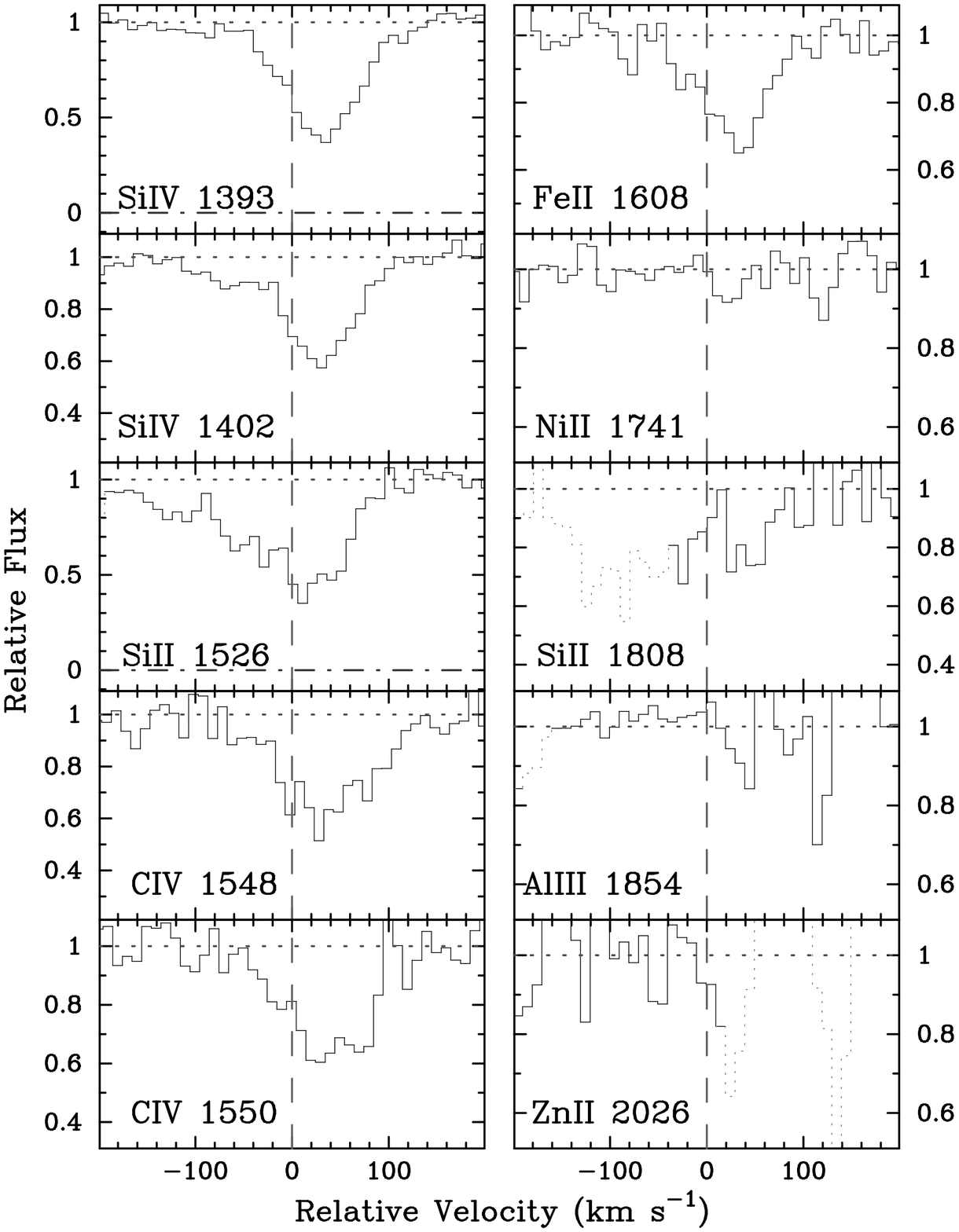}
\figcaption{Velocity plot of the metal-line transitions for the 
damped \lya system at $z = 3.554$ toward PSS1802+5616.
The vertical line at $v=0$ corresponds to $z = 3.5539$.  
\label{fig:pss1802B_mtl}}
\end{center}
\end{figure}

\begin{figure}[ht]
\begin{center}
\includegraphics[height=6.1in, width=3.9in]{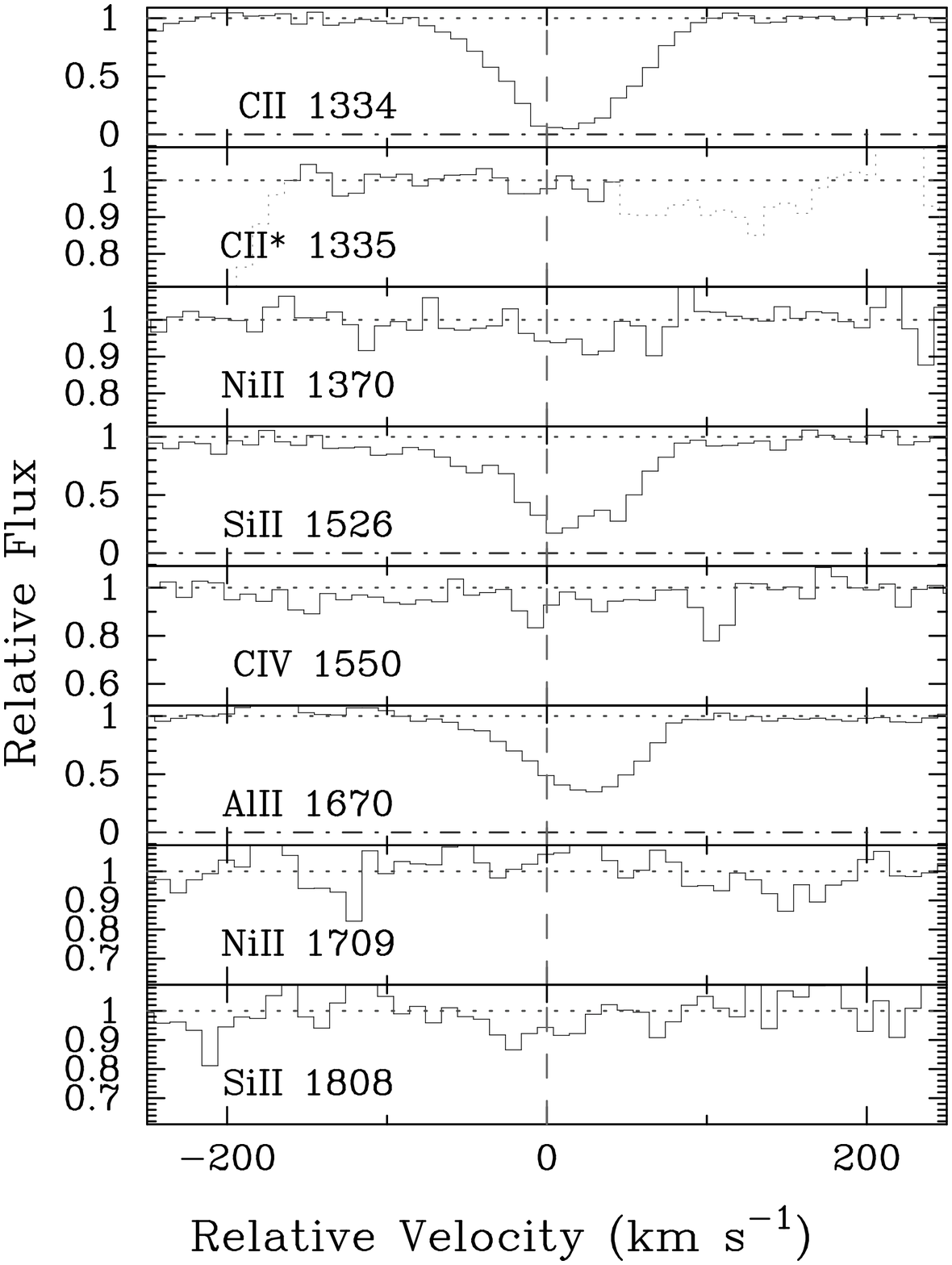}
\figcaption{Velocity plot of the metal-line transitions for the 
damped \lya system at $z = 3.762$ toward PSS1802+5616.
The vertical line at $v=0$ corresponds to $z = 3.7617$.  
\label{fig:pss1802C_mtl}}
\end{center}
\end{figure}

\begin{figure}[ht]
\begin{center}
\includegraphics[height=6.1in, width=3.9in]{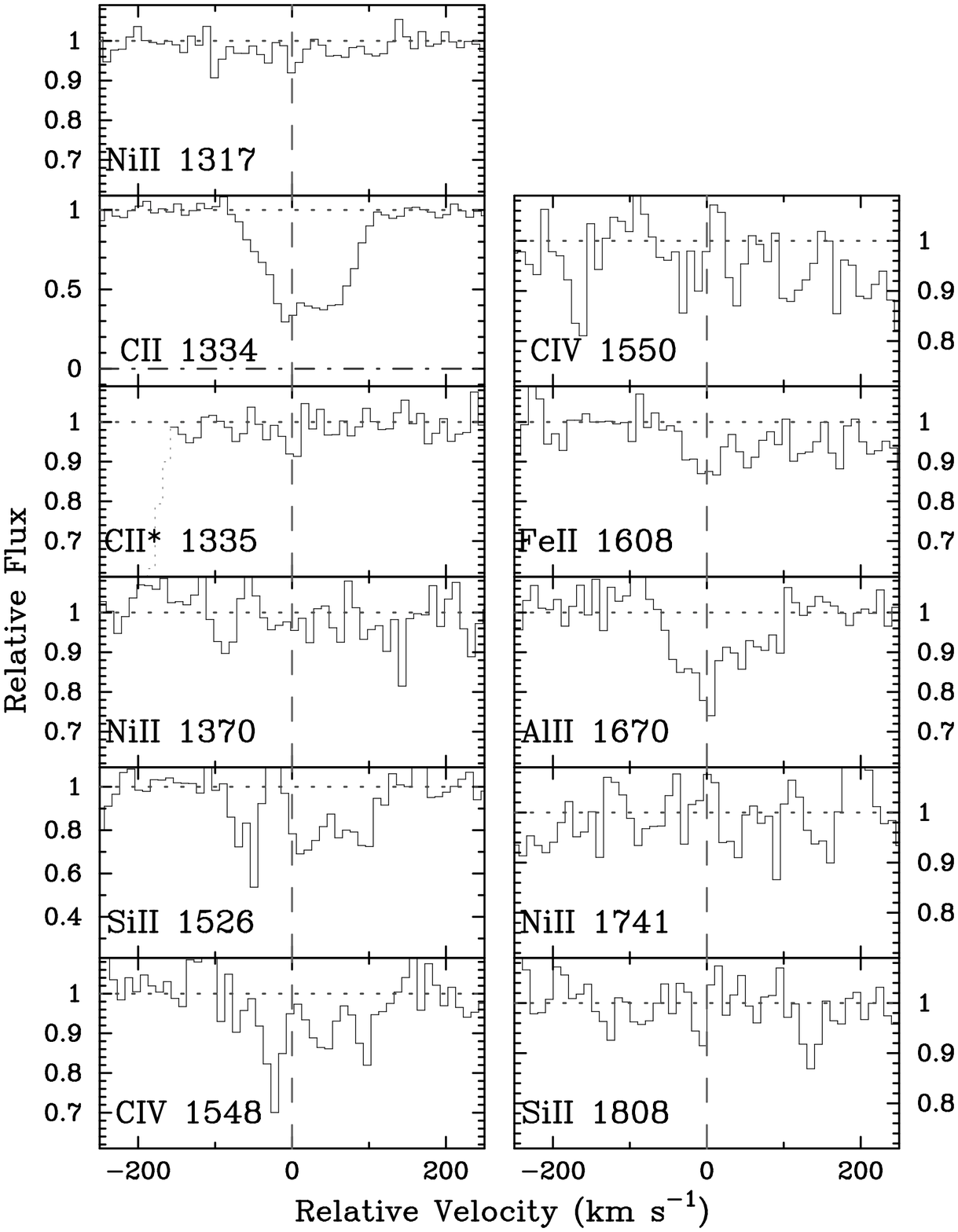}
\figcaption{Velocity plot of the metal-line transitions for the 
damped \lya system at $z = 3.811$ toward PSS1802+5616.
The vertical line at $v=0$ corresponds to $z = 3.8109$.  
\label{fig:pss1802D_mtl}}
\end{center}
\end{figure}

\begin{table}[ht]\footnotesize
\begin{center}
\caption{ {\sc
IONIC COLUMN DENSITIES: PSS1802+5616, $z = 3.391$ \label{tab:PSS1802+5616_3.391}}}
\begin{tabular}{lcccc}
\tableline
\tableline
Ion & $\lambda$ & AODM & $N_{\rm adopt}$ & [X/H] \\
\tableline
C  IV &1548.2&$13.990 \pm  0.026$\\  
C  IV &1550.8&$14.115 \pm  0.037$\\  
Al II &1670.8&$>13.152$&$>13.152$&$>-1.638$\\  
Al III&1854.7&$<12.441$\\  
Si II &1526.7&$>14.314$&$>14.314$&$>-1.546$\\  
Si II &1808.0&$<14.556$\\  
Fe II &1608.5&$14.256 \pm  0.037$&$14.256 \pm  0.037$&$-1.544 \pm  0.107$\\  
Ni II &1709.6&$<13.654$&$<13.654$&$<-0.896$\\  
Ni II &1751.9&$<13.717$\\  
Zn II &2026.1&$<12.405$&$<12.405$&$<-0.565$\\  
\tableline
\end{tabular}
\end{center}
\end{table}
\begin{table}[ht]\footnotesize
\begin{center}
\caption{ {\sc
IONIC COLUMN DENSITIES: PSS1802+5616, $z = 3.554$ \label{tab:PSS1802+5616_3.554}}}
\begin{tabular}{lcccc}
\tableline
\tableline
Ion & $\lambda$ & AODM & $N_{\rm adopt}$ & [X/H] \\
\tableline
C  IV &1548.2&$13.820 \pm  0.036$\\  
C  IV &1550.8&$14.063 \pm  0.039$\\  
Al III&1854.7&$<12.496$\\  
Si II &1526.7&$>14.239$&$>14.239$&$>-1.821$\\  
Si II &1808.0&$<15.122$\\  
Si IV &1393.8&$13.587 \pm  0.012$\\  
Si IV &1402.8&$13.664 \pm  0.020$\\  
Fe II &1608.5&$14.075 \pm  0.061$&$14.075 \pm  0.061$&$-1.925 \pm  0.117$\\  
Ni II &1741.6&$<13.299$&$<13.299$&$<-1.451$\\  
Zn II &2026.1&$<12.632$&$<12.632$&$<-0.538$\\  
\tableline
\end{tabular}
\end{center}
\end{table}
\begin{table}[ht]\footnotesize
\begin{center}
\caption{ {\sc
IONIC COLUMN DENSITIES: PSS1802+5616, $z = 3.762$ \label{tab:PSS1802+5616_3.762}}}
\begin{tabular}{lcccc}
\tableline
\tableline
Ion & $\lambda$ & AODM & $N_{\rm adopt}$ & [X/H] \\
\tableline
C  II &1334.5&$>14.674$&$>14.674$&$>-2.466$\\  
C  II*&1335.7&$<12.853$\\  
C  IV &1550.8&$<13.374$\\  
Al II &1670.8&$>12.960$&$>12.960$&$>-2.080$\\  
Si II &1526.7&$>14.370$&$>14.370$&$>-1.740$\\  
Si II &1808.0&$<14.747$\\  
Ni II &1370.1&$<13.332$&$<13.332$&$<-1.468$\\  
Ni II &1709.6&$<13.620$\\  
\tableline
\end{tabular}
\end{center}
\end{table}
\begin{table}[ht]\footnotesize
\begin{center}
\caption{ {\sc
IONIC COLUMN DENSITIES: PSS1802+5616, $z = 3.811$ \label{tab:PSS1802+5616_3.811}}}
\begin{tabular}{lcccc}
\tableline
\tableline
Ion & $\lambda$ & AODM & $N_{\rm adopt}$ & [X/H] \\
\tableline
C  II &1334.5&$>14.412$&$>14.412$&$>-2.528$\\  
C  II*&1335.7&$<12.978$\\  
C  IV &1548.2&$13.407 \pm  0.079$\\  
C  IV &1550.8&$<13.393$\\  
Al II &1670.8&$12.393 \pm  0.045$&$12.393 \pm  0.045$&$-2.447 \pm  0.205$\\  
Si II &1526.7&$13.870 \pm  0.102$&$13.870 \pm  0.102$&$-2.040 \pm  0.225$\\  
Si II &1808.0&$<14.808$\\  
Fe II &1608.5&$13.665 \pm  0.104$&$13.665 \pm  0.104$&$-2.185 \pm  0.225$\\  
Ni II &1317.2&$<13.078$&$<13.078$&$<-1.522$\\  
Ni II &1370.1&$<13.411$\\  
Ni II &1741.6&$<13.577$\\  
\tableline
\end{tabular}
\end{center}
\end{table}

\break

\subsection{PSS1535+2943, $z = 3.202,3.761$ \label{subsec:PSS1535+2943}}

The \lya profiles for the two damped systems along this sightline
are displayed in Figures~\ref{fig:pss1535A_lya} and \ref{fig:pss1535B_lya}
and they provide reasonably good measurements of $\N{HI}$ for each system.
The damped system at $z=3.202$ exhibits only a few metal-line transitions
outside the \lya forest and the majority of these are saturated
(Figure~\ref{fig:pss1535A_mtl}).  
Nevertheless, a reasonable estimate of the Fe metallicity can be
determined from the competing upper and lower limits (Fe\,II 1608,
1611; Table~\ref{tab:PSS1535+2943_3.202}).  
Regarding the DLA at $z=3.761$, its Fe\,II transitions are lost
in the thick band of atmospheric absorption at $\lambda \approx 7650$\AA\
but the system exhibits unsaturated Si\,II and Al\,II profiles which provide
accurate metal abundances (Figure~\ref{fig:pss1535B_mtl}, 
Table~\ref{tab:PSS1535+2943_3.761}).

\subsection{PSS1715+3809, $z = 3.341$ \label{subsec:PSS1715+3809_3.341}}
This is the only quasar reported here with a single damped \lya system.
Its \lya profile and corresponding metal-lines are presented in 
Figures~\ref{fig:pss1715_lya} and \ref{fig:pss1715_mtl}.  This damped
system  is well separated from the emission redshift of the quasar and
therefore exhibits few metal-line transitions outside the \lya forest.
Nevertheless, the Fe\,II 1608 and Al\,II 1670 profiles indicate it 
possesses one of the lowest metallicities to date.
Table~\ref{tab:PSS1715+3809_3.341} lists the values and limits measured
for this DLA.

\subsection{PSS1802+5616, $z = 3.391,3.554,3.762,3.811$ \label{subsec:PSS1802+5616}}
This sightline marks the first quasar with four foreground damped \lya
systems.  The \lya profiles are plotted in 
Figures~\ref{fig:pss1802A_lya}-\ref{fig:pss1802C_lya} and with the
exception of the pair of DLA at $z \sim 3.8$, the $\N{HI}$ values are well constrained.
The proximity of the pair and the poor sampling of the core of the \lya profiles
yield less constrained H\,I column densities.
The metal-line transitions for the four DLA are presented in 
Figures~\ref{fig:pss1802A_mtl}-\ref{fig:pss1802D_mtl}.  
The systems exhibit rather similar metallicity, [M/H]~$\approx -1.8$, and are
otherwise unremarkable
(Tables~\ref{tab:PSS1802+5616_3.391}-\ref{tab:PSS1802+5616_3.811}).

\begin{figure*}[ht]
\begin{center}
\includegraphics[height=5.5in, width=4.0in,angle=90]{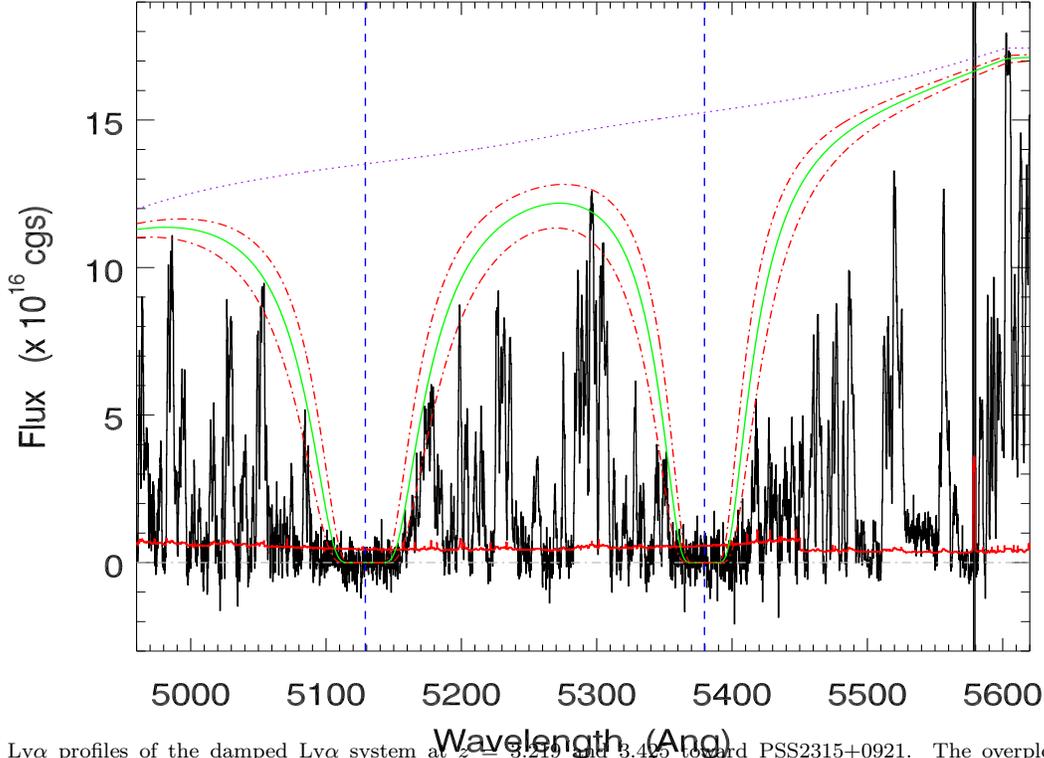}
\figcaption{\lya profiles of the damped \lya system at $z=3.219$
and 3.425 toward PSS2315+0921.
The overplotted solid line and accompanying
dash-dot lines trace the best fit solution corresponding to 
$\log \N{HI} = 21.35^{+0.15}_{-0.15}$ and 
21.10$^{+0.20}_{-0.20}$ respectively.
The proximity of the two DLA complicates the solution and the
quasar continuum is rapidly changing across this spectrum in part
due to a higher redshift Lyman limit system.
Finally, the metal-lines for the $z=3.425$ DLA are very extended
in velocity space, further complicating the $\N{HI}$ fit.  Here
we have assumed a single component which we found gives a comparable
total $\N{HI}$ as a multi-component fit.  Nevertheless, we attribute
a 0.2~dex uncertainty to this measurement.
\label{fig:pss2315_lya}}
\end{center}
\end{figure*}

\begin{figure}[ht]
\begin{center}
\includegraphics[height=6.1in, width=3.9in]{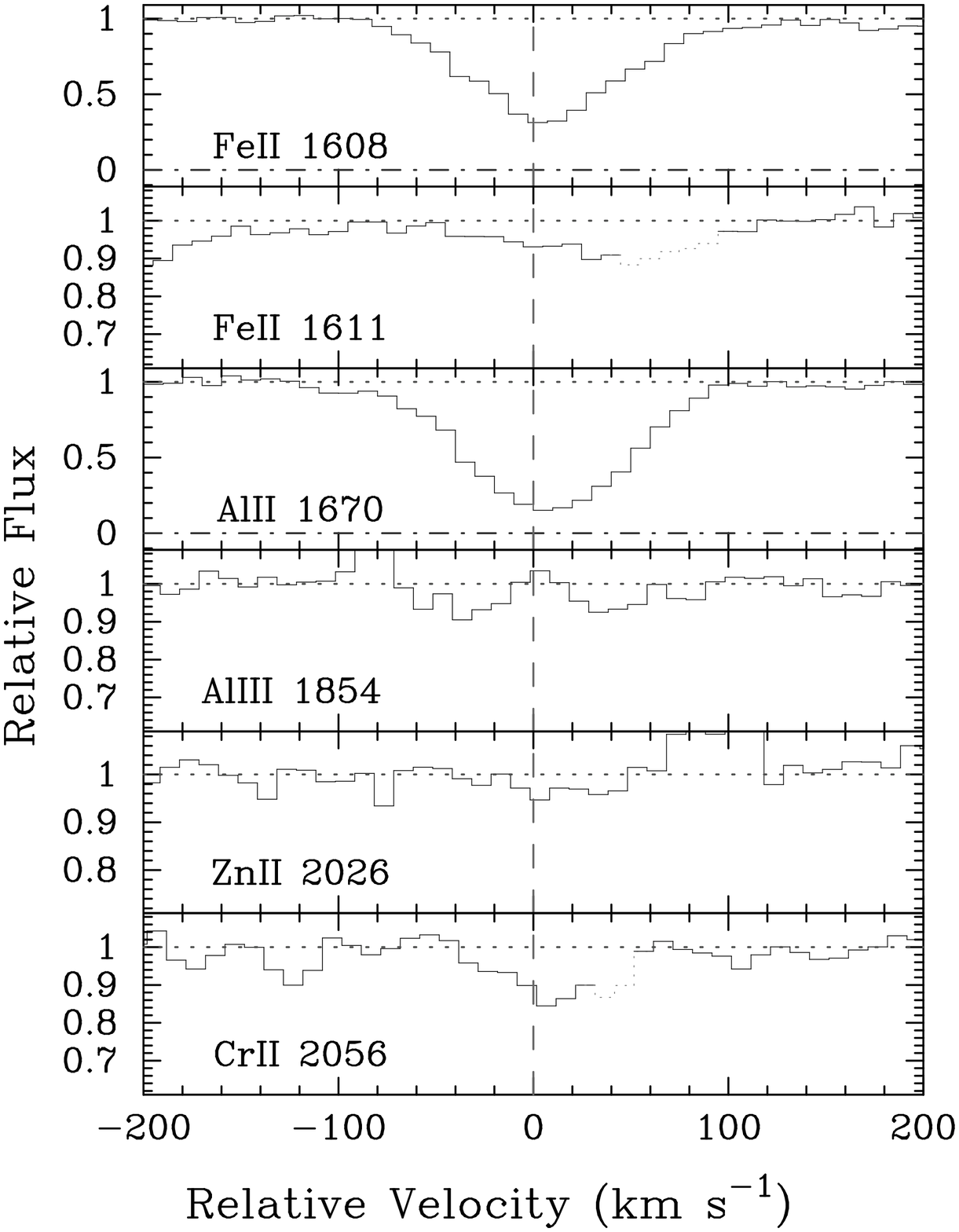}
\figcaption{Velocity plot of the metal-line transitions for the 
damped \lya system at $z = 3.219$ toward PSS2315+0921.
The vertical line at $v=0$ corresponds to $z = 3.2191$.  
\label{fig:pss2315A_mtl}}
\end{center}
\end{figure}

\begin{figure}[ht]
\begin{center}
\includegraphics[height=6.1in, width=3.9in]{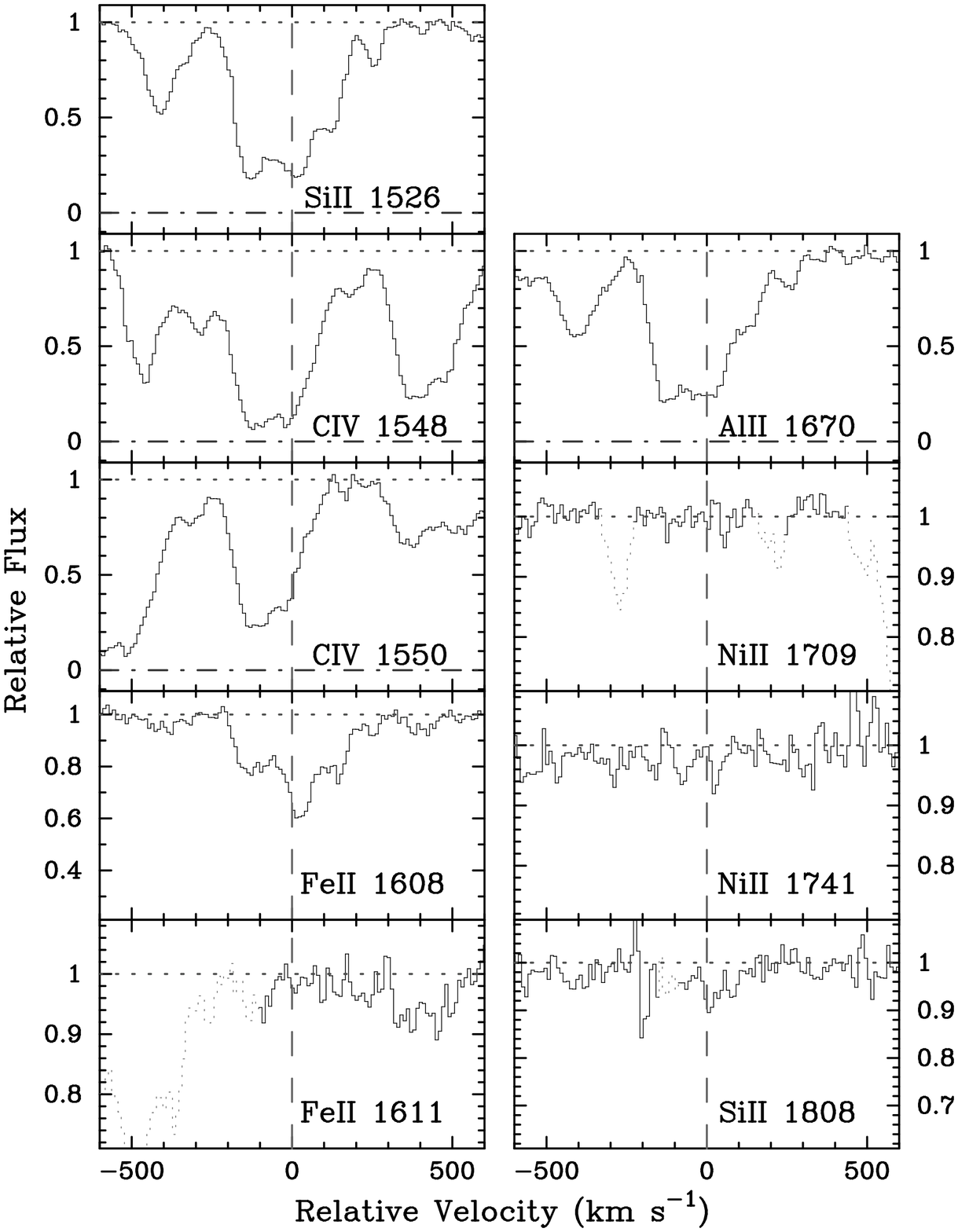}
\figcaption{Velocity plot of the metal-line transitions for the 
damped \lya system at $z = 3.425$ toward PSS2315+0921.
The vertical line at $v=0$ corresponds to $z = 3.4252$.  
\label{fig:pss2315B_mtl}}
\end{center}
\end{figure}

\begin{table}[ht]\footnotesize
\begin{center}
\caption{ {\sc
IONIC COLUMN DENSITIES: PSS2315+0921, $z = 3.219$ \label{tab:PSS2315+0921_3.219}}}
\begin{tabular}{lcccc}
\tableline
\tableline
Ion & $\lambda$ & AODM & $N_{\rm adopt}$ & [X/H] \\
\tableline
Al II &1670.8&$>13.240$&$>13.240$&$>-2.600$\\  
Al III&1854.7&$<12.284$\\  
Cr II &2056.3&$<13.191$&$<13.191$&$<-1.829$\\  
Fe II &1608.5&$>14.557$&$>14.557$&$>-2.293$\\  
Fe II &1611.2&$<14.983$\\  
Zn II &2026.1&$<11.947$&$<11.947$&$<-2.073$\\  
\tableline
\end{tabular}
\end{center}
\end{table}
\begin{table}[ht]\footnotesize
\begin{center}
\caption{ {\sc
IONIC COLUMN DENSITIES: PSS2315+0921, $z = 3.425$ \label{tab:PSS2315+0921_3.425}}}
\begin{tabular}{lcccc}
\tableline
\tableline
Ion & $\lambda$ & AODM & $N_{\rm adopt}$ & [X/H] \\
\tableline
C  IV &1548.2&$>15.087$\\  
C  IV &1550.8&$>14.989$\\  
Al II &1670.8&$>13.781$&$>13.781$&$>-1.809$\\  
Si II &1526.7&$>15.016$&$15.149 \pm  0.049$&$-1.511 \pm  0.206$\\  
Si II &1808.0&$15.149 \pm  0.049$\\  
Fe II &1608.5&$>14.635$&$>14.635$&$>-1.965$\\  
Fe II &1611.2&$<14.979$\\  
Ni II &1709.6&$<13.359$&$<13.359$&$<-1.991$\\  
Ni II &1741.6&$<13.491$\\  
\tableline
\end{tabular}
\end{center}
\end{table}
\subsection{PSS2315+0921, $z = 3.219,3.425$ \label{subsec:PSS2315+0921}}

The two damped \lya systems along this sightline exhibit 
H\,I column densities in excess of $10^{21} \cm{-2}$
(Figure~\ref{fig:pss2315_lya}) and the largest in this sample.  
Examining the metal-line transitions for the two DLA
(Figures~\ref{fig:pss2315A_mtl},\ref{fig:pss2315B_mtl})
one notes significant differences.  Specifically, the DLA at $z=3.219$
is relatively metal-poor and shows very simple kinematics.  In
contrast, the DLA at $z=3.425$ has metal-profiles extending
$\approx 1000 \mkms$, one of the largest velocity fields exhibited
by a DLA to date.  Even a 90$\%$ optical-depth determined velocity
width \citep{pw97} would exceed 400 \kms.
Tables~\ref{tab:PSS2315+0921_3.219} and \ref{tab:PSS2315+0921_3.425} 
list the column densities for these two galaxies.

\clearpage

\begin{sidewaystable*}\footnotesize
\begin{center}
\caption{ {\sc ABUNDANCE SUMMARY \label{tab:XHsum}}}
\begin{tabular}{lccrrrrrrrrrrrrrr}
\tableline
\tableline \tskip
Name & $z_{abs}$ & $\N{HI}$
& [C/H] & [O/H] & [Al/H] & [Si/H] & [Cr/H] & [Fe/H] & [Ni/H] & [Zn/H]\\
\tableline\tskip
PSS0007+2417&3.496&21.10&&&$>-2.341$&$-1.583$&$<-0.574$&$>-1.970$&$-1.821$&$<-1.380$\\  
PSS0007+2417&3.705&20.55&$>-2.610$&$>-2.435$&$>-2.145$&$>-1.738$&&$>-1.839$&$<-1.603$&\\  
PSS0007+2417&3.838&20.85&$>-3.046$&$>-2.947$&$-2.641$&$>-2.404$&&$-2.444$&$<-2.058$&\\  
PSS1535+2943&3.202&20.65&&&$>-1.580$&$>-1.468$&$<-1.131$&$>-1.547$&$<-0.928$&$<-0.789$\\  
PSS1535+2943&3.761&20.40&$>-2.621$&$>-2.429$&$-2.327$&$-2.020$&&&$<-1.197$&$< 0.062$\\  
PSS1715+3809&3.341&21.05&&&$-2.989$&$<-2.122$&$<-1.947$&$-2.807$&$<-2.006$&$<-1.607$\\  
PSS1802+5616&3.391&20.30&&&$>-1.638$&$>-1.546$&&$-1.544$&$<-0.896$&$<-0.565$\\  
PSS1802+5616&3.554&20.50&&&&$>-1.821$&&$-1.925$&$<-1.451$&$<-0.538$\\  
PSS1802+5616&3.762&20.55&$>-2.466$&&$>-2.080$&$>-1.740$&&&$<-1.468$&\\  
PSS1802+5616&3.811&20.35&$>-2.528$&&$-2.447$&$-2.040$&&$-2.185$&$<-1.522$&\\  
PSS2315+0921&3.219&21.35&&&$>-2.600$&&$<-1.829$&$>-2.293$&&$<-2.073$\\  
PSS2315+0921&3.425&21.10&&&$>-1.809$&$-1.511$&&$>-1.965$&$<-1.991$&\\  
\tskip \tableline
\end{tabular}
\end{center}
\end{sidewaystable*}

\begin{sidewaystable*}\footnotesize
\begin{center}
\caption{ 
{\sc RELATIVE ABUNDANCE SUMMARY \label{tab:XFesum}}}
\begin{tabular}{lccrrrrrrrrrrrrrr}
\tableline
\tableline \tskip
Name & $z_{abs}$ & $\N{HI}$
& [C/Fe] & [O/Fe] & [Al/Fe] & [Si/Fe] & [Cr/Fe] & [Ni/Fe] & [Zn/Fe]\\
\tableline\tskip
PSS0007+2417$^a$&3.496&21.10&&&$>-0.520$&$+ 0.238$&$<+ 1.247$&&$<+ 0.441$\\  
PSS0007+2417&3.705&20.55\\  
PSS0007+2417&3.838&20.85&$>-0.602$&$>-0.503$&$-0.197$&$>+ 0.040$&&$<+ 0.386$&\\  
PSS1535+2943&3.202&20.65\\  
PSS1535+2943$^b$&3.761&20.40&$>-0.294$&$>-0.102$&&$+ 0.307$&&$<+ 1.130$&$<+ 2.389$\\  
PSS1715+3809&3.341&21.05&&&$-0.182$&$<+ 0.685$&$<+ 0.860$&$<+ 0.801$&$<+ 1.200$\\  
PSS1802+5616&3.391&20.30&&&$>-0.094$&$>-0.002$&&$<+ 0.648$&$<+ 0.979$\\  
PSS1802+5616&3.554&20.50&&&&$>+ 0.104$&&$<+ 0.474$&$<+ 1.387$\\  
PSS1802+5616&3.762&20.55\\  
PSS1802+5616&3.811&20.35&$>-0.343$&&$-0.262$&$+ 0.145$&&$<+ 0.663$&\\  
PSS2315+0921&3.219&21.35\\  
PSS2315+0921&3.425&21.10\\  
\tskip \tableline
\end{tabular}
\end{center}
$^a$Ni is serving as a proxy for Fe\\
$^b$Al is serving as a proxy for Fe
\end{sidewaystable*}

\clearpage

\section{SUMMARY}

Tables~\ref{tab:XHsum} and \ref{tab:XFesum} present a summary of the
absolute and relative abundances of the \ndla\ damped \lya systems
in this dataset.  In Table~\ref{tab:XFesum}, where
we present abundances relative to Fe, we have 
considered Ni or Al as a proxy for Fe in a few cases as noted.

We have presented ionic column density measurements for the \ndla\
new damped \lya systems drawn from the PSS survey.
We have measured
metal-line column densities with the apparent optical depth method and
$\N{HI}$ values through qualitative fits to the \lya profiles.  
All of the data has been reduced and analysed with the same techniques.
Visit {\tt http://www.ucolick.org/$\sim$xavier/ESI} 
for tables, figures and updated measurements, 
{\tt http://www.astro.caltech.edu/$\sim$george/z4.qsos} for information
on the PSS quasars,
and {\tt http://kingpin.ucsd.edu/$\sim$hiresdla/} for the
HIRES damped \lya abundance database and relevant atomic data.  

\acknowledgments

The authors wish to recognize and acknowledge the very significant cultural
role and reverence that the summit of Mauna Kea has always had within the
indigenous Hawaiian community.  We are most fortunate to have the
opportunity to conduct observations from this mountain.
We acknowledge the Keck support staff for their efforts
in performing these observations.


\begin{thebibliography}{}

\bibitem[Castro, Djorgovski, \& de Carvalho (2003)]{castro03}  	
Castro, S., Djorgovski, S.G., \& de Carvalho, R. 2003, ApJS, submitted

\bibitem[Djorgovski et al.\ (1998)]{djg98}		
Djorgovski, S.G., Gal, R.R., Odewahn, S.C., de Carvalho, R.R.,
Brunner, R., Longo, G., \& Scaramella, R. 1998, in ``Wide Field
Surveys in Cosmology'', eds.\ S. Colombi \& Y. Mellier, 
Gif sur Yvette: Editions Frontie\`eres, p. 89 (astro-ph/9809187)

\bibitem[Djorgovski et al.\ (2001)]{djg01}		
Djorgovski, S.G., Mahabal, A., Brunner, R., Gal, R., 
Castro, S., de Carvalho, R.,
\& Odewahn, S.C. 2001, in: Virtual Observatories of the Future, eds.
R. Brunner, S.G. Djorgovski \& A. Szalay, ASPCS, 225, 52
(astro-ph/0012453)


\bibitem[Grevesse et al.\ (1996)]{grvss96}	
Grevesse, N., Noels, A., \& Sauval, A.J. 1996, In: Cosmic Abundances,
S. Holt and G. Sonneborn (eds.), ASPCS, V. 99, (BookCrafters: San Fransisco),
p. 117

\bibitem[Holweger (2001)]{holweger01}   
Holweger, H. 2001, in Solar and Galactic Composition, ed.\ R.F.
Wimmer-Schweingruber, (Berlin: Springer), 23

\bibitem[Kennefick et al.\ (1995a)]{kennefick95a}
Kennefick, J.D., de Carvalho, R., Djorgovski, S.G., Wilber, M., Dickson, E.,
Weir, N., Fayyad, U., \& Roden, J. 1995a, AJ, 110, 78

\bibitem[Kennefick et al.\ (1995b)]{kennefick95b}
Kennefick, J.D., Djorgovski, S.G., \& de Carvalho, R. 1995b, AJ, 110, 2553

\bibitem[Prochaska, Gawiser, \& Wolfe (2001)]{pgw01}  
Prochaska, J.X., Gawiser, E., \& Wolfe, A.M. 2001, \apj, 552, 99 (PGW01)

\bibitem[Prochaska et al.\ (2003)]{pro03}  		
Prochaska, J.X., Gawiser, E., Wolfe, A.M., Cooke, J.,
\& Gelino, D. 2003, \apjs, in press (P03)

\bibitem[Prochaska \& Wolfe (1997)]{pw97}  		
Prochaska, J. X. \& Wolfe, A. M. 1997, \apj, 486, 73

\bibitem[Savage and Sembach (1991)]{sav91}
Savage, B. D. and Sembach, K. R. 1991, \apj, 379, 245

\bibitem[Sheinis et al.\ (2002)]{sheinis02}		
Sheinis, A.I., Miller, J., Bigelow, B., Bolte, M., Epps, H.,
Kibrick, R., Radovan, M., \& Sutin, B. 2002, \pasp, 114, 851

\bibitem[Storrie-Lombardi and Wolfe (2000)]{storrie00}  
Storrie-Lombardi, L.J. \& Wolfe, A.M. 2000, \apj, 543, 552

\end{thebibliography}
\end{document}